\documentclass[a4paper]{article}

\usepackage{INTERSPEECH2020}
\usepackage{graphics}
\usepackage{cite}
\usepackage{authblk}
\usepackage{xcolor}
\usepackage{multirow}
\title{Converting Anyone's Emotion: \\Towards Speaker-Independent Emotional Voice Conversion}
\name{Kun Zhou $^ a$ \thanks{\textbf{Codes \& Speech samples:} \url{https://kunzhou9646.github.io/speaker-independent-emotional-vc/}}, Berrak Sisman $^{ b, a}$, Mingyang Zhang $^a$, Haizhou Li $^ a$}

\address{
  $^a$Department of Electrical and Computer Engineering, National University of Singapore, Singapore \\
  $^b$Singapore University of Technology and Design, Singapore}
\email{zhoukun@u.nus.edu, berraksisman@u.nus.edu, elezmin@nus.edu.sg, haizhou.li@nus.edu.sg}

\begin{document}

\maketitle
\begin{abstract}
Emotional voice conversion aims to convert the emotion of speech from one state to another while preserving the linguistic content and speaker identity. The prior studies on emotional voice conversion are mostly carried out under the assumption that emotion is speaker-dependent. We consider that there is a common code between speakers for emotional expression in a spoken language, therefore, a speaker-independent mapping between emotional states is possible. In this paper, we propose a speaker-independent emotional voice conversion framework, that can convert anyone's emotion without the need for parallel data. We propose a VAW-GAN based encoder-decoder structure to learn the spectrum and prosody mapping. We perform prosody conversion by using continuous wavelet transform (CWT) to model the temporal dependencies. We also investigate the use of F0 as an additional input to the decoder to improve emotion conversion performance. Experiments show that the proposed speaker-independent framework achieves competitive results for both seen and unseen speakers.

\end{abstract}

\vspace{-3mm}
\section{Introduction}
%
Emotional voice conversion (EVC) is a type of voice conversion (VC) that converts the emotional state of speech from one to another while preserving the linguistic content and speaker identity. It has various applications in expressive speech synthesis~\cite{liu2019teacher}, such as intelligent dialogue systems, voice assistants, and conversational agents. 

In general, voice conversion refers to the conversion of speaker identity while preserving the linguistic information. The earlier VC studies include Gaussian mixture model (GMM) \cite{toda2007voice}, partial least square regression (PLSR) \cite{helander2010voice}, NMF-based exemplar-based sparse representation  \cite{aihara2014exemplar, sisman2018voice} and group sparse representation \cite{ccicsman2017sparse}. Recent deep learning approaches, such as deep neural network (DNN) \cite{hinton2006fast}, and variational autoencoder (VAE) \cite{huang2019investigation,qian2019zero,Qian_2020} have greatly improved voice conversion quality. 

Spectral mapping has been the main focus of conventional voice conversion; however, prosody mapping has not been given the same level of attention. We note that emotion is inherently supra-segmental and hierarchical in nature, that is manifested both in the spectrum and prosody \cite{xu2011speech,latorre2008multilevel}. Therefore, it is insufficient for emotional voice conversion to just convert the spectral features frame-by-frame.



Statistical modelling for prosody conversion represents one of the successful attempts. In \cite{tao2006prosody}, the pitch contour was decomposed into a hierarchical structure with a classification-regression tree, then converted by GMM and regression-based clustering methods. A GMM-based model \cite{aihara2012gmm} was proposed to handle both spectrum and prosody conversion. 
Another strategy is to create a source and target dictionary and estimate a sparse mapping using exemplar-based techniques with NMF \cite{aihara2014exemplar}. Moreover, an emotional voice conversion model combining hidden Markov model (HMM), GMM, and F0 (fundamental frequency) segment selection method was proposed in \cite{inanoglu2009data}, which can convert pitch, duration and spectrum. The prior studies serve as a source of inspiration for this work. 

There have been studies on deep learning approaches for emotional voice conversion with parallel training data, such as deep neural network \cite{lorenzo2018investigating}, deep belief network \cite{luo2016emotional} and deep bi-directional long-short-term memory network \cite{ming2016deep}. More recently, other methods, such as sequence-to-sequence model \cite{robinson2019sequence} and rule-based model \cite{xue2018voice}, were also proven to be effective. To eliminate the need for parallel training data, autoencoders \cite{gao2018nonparallel} and cycle-consistent generative adversarial networks (CycleGAN) \cite{zhou2020transforming} based emotional voice conversion frameworks were proposed and shown remarkable performance. We note that these frameworks are designed for a specific speaker; therefore, they are called speaker-dependent frameworks. 





It is believed that emotional expression and perception present individual variations influenced by personalities, languages and cultures \cite{kotti2012speaker, fersini2009audio,arnold1960emotion,dai2009comparing}. Simultaneously, they also share some common cues across individuals regardless of their identities and backgrounds \cite{arnold1960emotion,dai2009comparing, schuller2005speaker}. 
In the field of emotion recognition, speaker-independent emotion recognition demonstrates a more robust, stable and better generalization ability than the speaker-dependent ones \cite{kotti2012speaker}.
However, so far, few researchers have explored the speaker-independent emotional voice conversion. Most related studies, such as \cite{shankar2019multi}, have only dealt with a multi-speaker model at most. 
\begin{figure*}
    \centering
    \includegraphics[width=12cm]{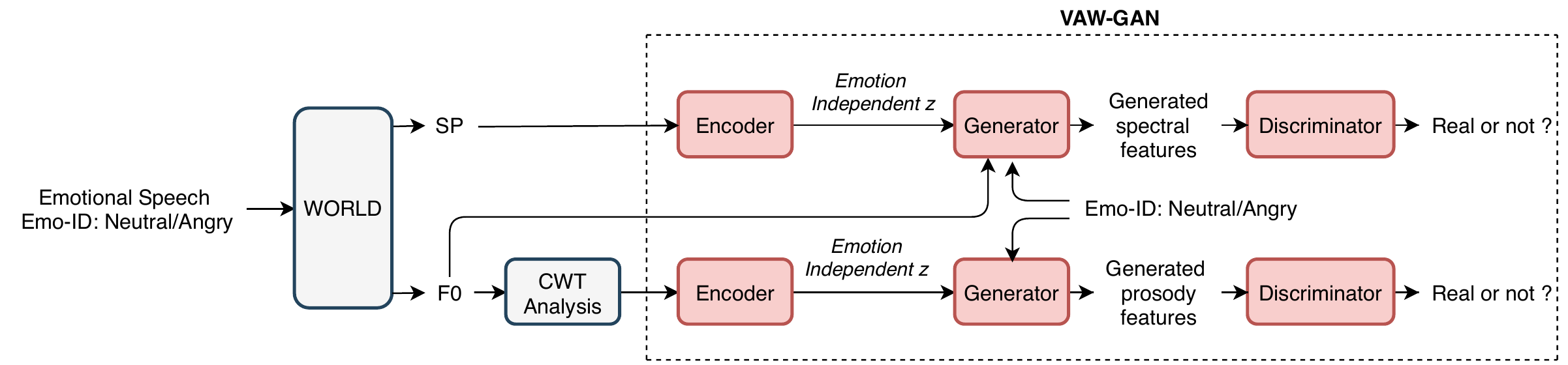}
    \vspace{-3mm}
    \caption{The training phase of the proposed VAW-GAN-based EVC. Red boxes are involved in the training, while grey boxes are not.}
    \label{fig:training}
    \vspace{-3mm}
\end{figure*}

CycleGAN is an effective solution for voice conversion without parallel training data; however, it is more suitable for pair-wise conversion. An encoder-decoder structure, such as variational autoencoding Wasserstein generative adversarial network (VAW-GAN) \cite{hsu2017voice} would be more suitable to learn the emotion-independent representations. 
The main contributions of this paper include: 1) we study emotion through speaker-independent perspective for voice conversion; 2) we propose a VAW-GAN architecture and its training framework does not require parallel training data; and 3) we study prosody modelling, and propose F0 conditioning for emotion-independent encoder training. 

The paper is organized as follows: In Section 2, we motivate the perspective of speaker-independent emotion. In Section 3, we introduce the proposed emotional voice conversion framework. In Section 4, we report the experiments. Section 5 concludes the study.

\section{Speaker-Independent Perspective on Emotion}
Speech is more than just words. It carries emotions of the speaker. Emotion reflects the intent, mood and temperament of the speaker, and plays an important role in decision making and opinion expression \cite{arnold1960emotion}. Emotion is also highly complex with multiple signal attributes concerning spectrum and prosody, which makes it difficult to disentangle and synthesize \cite{xu2011speech}.

Previous studies have revealed that basic emotions can be expressed and recognized through universal principles which are innately shared across human culture \cite{ekman1992argument}. In general, it is also commonly believed that emotional states in speech have an impact on the speech production mechanism across the glottal source and vocal tract of the individuals \cite{kane2014phonetic}. The studies prompt us  to investigate speech emotions from a speaker-independent perspective \cite{schuller2005speaker}. Studies have also shown possible ways of speaker-independent emotion representation for both seen and unseen speakers over a large multi-speakers emotional corpus \cite{kotti2012speaker}, emotion feature extraction and classifiers \cite{akccay2020speech}.


To validate the idea of speaker-independent emotion elements across speakers \cite{schuller2005speaker,kotti2012speaker}, we conduct a preliminary study using CycleGAN-based emotional voice conversion framework \cite{zhou2020transforming}, which is designed for speaker-dependent EVC. In this study, we train a network with two conversion pipelines for the mapping of spectrum and prosody (CWT-based F0 features) respectively.  
We train the network on one specific speaker and test it for both the intended (seen) and unseen speaker. In Table \ref{tab:cyclegan}, we report the performance of spectrum conversion in terms of Mel-cepstral distortion (MCD) and log-spectral distortion (LSD) \cite{huang2019investigation,qian2019zero,Qian_2020, sisman2019group}; and that of prosody conversion in terms of Pearson correlation coefficient (PCC) and root mean square error (RMSE) of F0 contours \cite{sisman2019group}. Zero effort represents the cases where we directly compare the speech of source and target emotions without any conversion.

We observe that the speaker-dependent CycleGAN system performs for the unseen speaker pretty well, which is encouraging. As shown in Table \ref{tab:cyclegan}, the results for the unseen speaker are clearly better than those for Zero Effort in terms of MCD and LSD for spectrum, and PCC and RMSE for prosody, despite the fact that it \cite{zhou2020transforming} does not have any information about the unseen speaker in advance. Encouraged by this observation, we propose a speaker-independent emotional voice conversion framework that converts anyone's emotion.
\begin{table}[h]
\vspace{-1mm}
\centering
\caption{ MCD {[}dB{]}, LSD {[}dB{]}, PCC and RMSE {[}Hz{]} results for seen and unseen speakers of CycleGAN-based EVC \cite{zhou2020transforming}.}
\begin{tabular}{l c  c  c c}
\hline
\textbf{Speaker} & \textbf{MCD} & \textbf{LSD} & \textbf{PCC} & \textbf{RMSE}  \\ \hline
Seen   &  4.948       & 7.028 & 0.721 &54.043\\ 
Unseen  &  5.131      & 7.298  &0.594 &62.826\\ 
Seen (Zero effort)  &  5.210      & 7.383  &0.571 &62.242\\
Unseen (Zero effort)   & 5.296 & 7.400  & 0.440 & 66.646\\ \hline
\end{tabular}
\label{tab:cyclegan}

\end{table}

\section{Speaker-Independent EVC}

\begin{figure*}
    \centering
    \includegraphics[width=14cm]{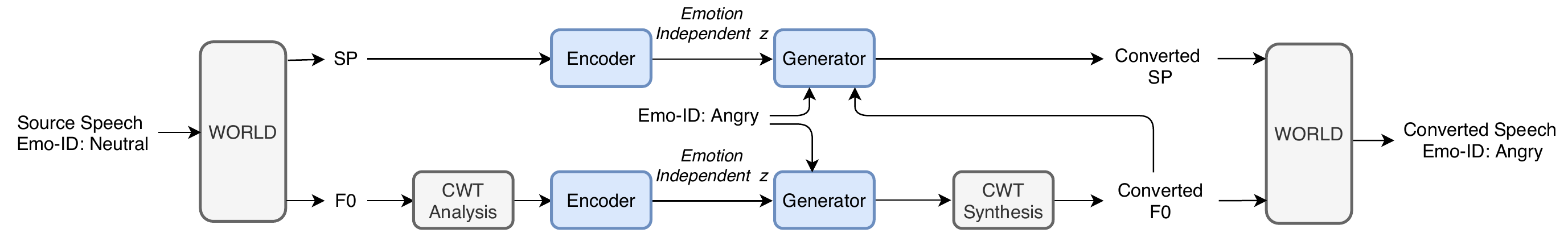}
    \vspace{-2mm}
    \caption{The run-time conversion phase of the proposed VAW-GAN-based EVC. Blue boxes represent the trained networks.}
    \label{fig:conversion}
    \vspace{-5mm}
\end{figure*}

An encoder-decoder structure, such as VAW-GAN \cite{hsu2017voice}, allows us to learn the emotion-independent representations using the encoder.  Instead of CycleGAN, we propose to take advantage of the encoder-decoder structure of VAW-GAN to formulate a speaker-independent emotional voice conversion framework.

 We first extract spectral (SP) and F0 features using WORLD vocoder. As F0 contains segmental information, it is insufficient to use a Logarithm Gaussian (LG)-based linear transformation to convert F0 contour~\cite{sisman2019group, csicsman2017transformation,luo2016emotional}. We perform CWT decomposition of F0 to describe the prosody from  micro-prosody level to utterance level, in a similar way reported in \cite{suni2013wavelets}. We believe that  speaker-dependent and independent components~\cite{sisman2018wavelet} in F0 contour should be dealt with differently. CWT decomposition of F0 allows the encoder to learn the speaker-independent emotion pattern between two speakers~\cite{csicsman2017transformation}.  As CWT is sensitive to the discontinuities in F0, the following preprocessing steps are needed: 1) linear interpolation over unvoiced regions, 2) transformation of  F0 from linear to a logarithmic scale, and  3)  normalization of the resulting F0 to zero mean and unit variance \cite{zhou2020transforming, sisman2019group, csicsman2017transformation}. 
\vspace{-2mm}
\subsection{Training}
\vspace{-1mm}
The proposed VAW-GAN and its training procedure are shown in Figure \ref{fig:training}. It was found that separate training of spectrum and prosody achieves better performance than joint training for emotion conversion \cite{zhou2020transforming}. Therefore, we propose to train two networks separately: 1) a VAW-GAN model conditioned on F0 for spectrum conversion, denoted as \textit{VAW-GAN for Spectrum}, and 2) a VAW-GAN model for prosody conversion denoted as \textit{VAW-GAN for Prosody}. Each network consists of three components, that are 1) encoder, 2) generator/decoder, and 3) discriminator.

During the training of \textit{VAW-GAN for Spectrum} and \textit{VAW-GAN for Prosody}, the encoder is exposed to input frames from multiple speakers with different emotions. The encoder learns the emotion-independent patterns among multiple speakers and transforms the input features into a latent code $z$. In this case, we assume that the latent code $z$ is emotion-independent and only contains the information of speaker identity and phonetic content. We use a one-hot vector as emotion ID to represent different emotions and provide the emotion information to the generator/decoder. 
Since the spectral features in \textit{VAW-GAN for Spectrum} training highly depend on F0 and contain prosodic information, we propose to use F0 as an additional condition to decoder, to disentangle the spectral features from the prosody, as shown in Figure \ref{fig:training}. 

 We then train the generative model based on adversarial learning for both spectrum and prosody, to find an optimal solution through a min-max game: the discriminator tries to maximize the loss between the real and reconstructed features, while the generator tries to minimize it \cite{ak2019attribute, ak2020semantically}. It allows us to achieve high-quality converted speech with target emotion that is defined by the emotion ID.
\vspace{-2mm}
\subsection{Run-time Conversion}
\vspace{-1mm}
The run-time conversion phase is given in Figure \ref{fig:conversion}. Similar to that of the training phase, for prosody conversion, we perform CWT on F0 to decompose F0 into coefficients of different time scales, that  are taken by prosody encoder to generate the converted F0 for a designated emotion ID. As for spectrum conversion, we propose to condition the generator/decoder on the converted CWT-based F0 coefficients together with emotion ID. Then, the converted spectral features are obtained through the trained \textit{VAW-GAN for Spectrum}. Finally, we use WORLD vocoder to synthesize the converted emotional speech. We note that aperiodicity (AP) is directly copied from the source speech. 
\vspace{-6mm}
\subsection{Effect of F0 Conditioning}
\vspace{-1mm}
We note that, for both spectrum and prosody conversion, the encoder is trained with input features from multiple speakers with different emotions to generate the emotion-independent latent code $z$. The decoder is conditioned on the target emotion ID to generate the speech. Since the spectral features are highly dependent on F0 features and also carry prosodic information, it is insufficient to train an emotion-independent encoder only using the one-hot emotion ID. Thus, we propose to add F0 as an additional input to the generator, that aims to force the encoder to learn only an  emotion-independent representation. F0 conditioning provides remarkable improvement over the baseline in both objective and subjective evaluation. 
\vspace{-2mm}
\subsection{Comparison with Related Work}
The proposed EVC framework is unique in the sense that it eliminates the need for: 1) parallel training data, 2) any alignment technique, 3) any speaker embedding, and 4) external modules such as speech recognizer. It shares a similar motivation with other VC frameworks based on conditional VAE \cite{huang2019investigation,Qian_2020} regarding F0 conditioning, but differs in many ways. For example, 1) we study emotion conversion, while \cite{huang2019investigation,Qian_2020} focus on speaker identity conversion; 2) through F0 conditioning mechanism, we eliminate the residual prosodic information in the latent code $z$ to make it emotion-independent, while \cite{huang2019investigation,Qian_2020} focus on the generation of speaker-independent latent code, and do not study the emotion perspective; 3) We propose to condition the generator/decoder on the converted CWT-based F0 features at run-time which has not been studied before. 

\vspace{-3mm}
\section{Experiments}

As a large emotional multi-speaker voice conversion dataset is not publicly available, we combine three different emotional speech corpora to conduct experiments, that are: 1) an English emotional speech corpus \cite{liu2014emotional}, 2) EmoV-DB \cite{adigwe2018emotional}, and 3) JL-Corpus \cite{james2018open}.  We train the networks using the speech data of three female speakers from the first two datasets and conduct emotion conversion on these three speakers and another two female speakers randomly chosen from JL-Corpus for evaluation. We call these two speakers from JL-Corpus as \textit{unseen speakers}, since the framework has no prior information of these speakers during training. Those involved in both training and conversion phase are denoted as \textit{seen speakers}. 

We choose two common emotions of these three datasets, that are 1) neutral and 2) angry. In all experiments, we conduct emotion conversion from neutral to angry. We conduct both objective and subjective experiments with 2 minutes of evaluation data to assess the system  performance in a comparative study.

\vspace{-3mm}
\subsection{Experimental Setup}
\vspace{-1mm}
As illustrated in Figure 1, we train two similar VAW-GAN pipelines for both spectrum and prosody conversion. The encoder for both frameworks is a 5-layer 1D convolutional neural network (CNN) with a kernel size of 7 and a stride of 3 followed by a fully connected layer. Its output channel is $\{16,32,64,128,256\}$. The latent code is 128-dimensional and assumed to have a standard normal distribution. 

In prosody conversion pipeline, the emotion ID is a 10-dimensional one-hot vector, that is concatenated with the latent code to generate a 138-dimensional vector and then merged by a fully connected layer. In spectrum conversion pipeline, a one-dimensional F0 is concatenated together with the latent code and emotion ID into a 139-dimensional vector. For GAN, the generator is a 4-layer 1D CNN with kernel sizes of $\{9,7,7,1025\}$ and strides of $\{3,3,3,1\}$, and the output channel is $\{32,16,8,1\}$. The discriminator is a 3-layer 1D CNN with kernel sizes of $\{7,7,115\}$ and a stride of $\{3,3,3\}$ followed by a fully connected layer. Its output channels are $\{16,32,64\}$. We train the networks by using RMSProp with a learning rate of 1e-5 and set the batch size as 256 for 45 epochs.
\vspace{-6mm}
\subsection{Objective Evaluation}
We perform objective evaluation to assess the performance of both spectrum and prosody conversion. We use MCD and LSD for spectrum conversion evaluation, while PCC is used for prosody conversion evaluation. In this section, the proposed VAW-GAN-based EVC framework given in Figure 1 is denoted as \textit{CWT-C-VAWGAN}. The baseline framework, denoted as \textit{C-VAWGAN}, converts spectrum with {VAW-GAN} conditioned on LG-based F0 without CWT decomposition, where F0 is converted in a traditional manner with LG-based linear transformation \cite{sisman2019group}. In Table \ref{tab: table}, we report comprehensive experimental results for both seen and unseen speakers.



\begin{table*}[ht]
\centering
\caption{ Average MCD {[}dB{]}, LSD {[}dB{]} and PCC of C-VAWGAN with LG-based F0 and CWT-C-VAWGAN. }
\vspace{-3mm}
\begin{tabular}{l||c c||c c||c c}
\hline
\multirow{2}{*}{Framework} & \multicolumn{2}{c||}{MCD {[}dB{]}}                  & \multicolumn{2}{c||}{LSD {[}dB{]}} & \multicolumn{2}{c}{PCC}     \\ \cline{2-7} 
                           & Seen & \multicolumn{1}{l||}{Unseen} & Seen    & Unseen   & Seen  & Unseen  \\ \hline
C-VAWGAN                   & 4.441        & 4.685                               & 6.188          & 6.286            & 0.750        & 0.630          \\ 
\textbf{CWT-C-VAWGAN}               & 4.439        & 4.683                               & 6.161          & 6.275            & 0.776        & 0.691          \\ \hline 
\end{tabular}
\vspace{-5mm}
\label{tab: table}
\end{table*}

We observe that the proposed \textit{CWT-C-VAWGAN} framework outperforms the baseline in terms of spectrum conversion by achieving consistently lower LSD and MCD values for both seen and unseen speakers. This shows that, in terms of F0 conditioning, CWT-based converted F0 features are more effective than the LG-based F0 features. We also note that \textit{CWT-C-VAWGAN} achieves comparable results between seen and unseen speakers. The results validate the idea of speaker-independent EVC in spectrum conversion.

We compare the \textit{CWT-C-VAWGAN} framework with the baseline in terms of prosody conversion. We note that  \textit{CWT-C-VAWGAN} consistently outperforms the baseline by achieving higher PCC for both seen and unseen speakers. Moreover, \textit{CWT-C-VAWGAN} reports a closer PCC between seen and unseen speaker than \textit{C-VAWGAN}.  These results validate the idea of speaker-independent EVC in prosody conversion.




\subsection{Subjective Evaluation}

We further conduct four listening experiments to assess \textit{CWT-C-VAWGAN} in terms of speech quality, emotion similarity and speaker similarity. 15 subjects participated in all experiments, each listening to 110 converted utterances. As a reference baseline,  \textit{CWT-VAWGAN} denotes the VAW-GAN system that converts spectrum and CWT-based F0 without conditioning the generator on F0.

We first report the mean opinion score (MOS) of the proposed \textit{CWT-C-VAWGAN} and baseline \textit{CWT-VAWGAN} for seen speakers. We note that both frameworks are based on \textit{VAW-GAN} for spectrum and prosody conversion, but \textit{CWT-C-VAWGAN} conditions the generator on additional CWT-based F0 features. As shown in Table \ref{tab:mos}, \textit{CWT-C-VAWGAN} outperforms \textit{CWT-VAWGAN} with a higher MOS score. The results confirm the effectiveness of the proposed CWT-based F0 conditioning. 

\begin{table}[ht]
\vspace{-3mm}
\centering
\caption{MOS results with 95\% confidence interval of CWT-VAWGAN and CWT-C-VAWGAN.}
\vspace{-2mm}
\begin{tabular}{c || c c}
\hline
Framework                                                           & MOS\\ \hline
\textit{CWT-VAWGAN}                                                          & 2.731 $\pm$ 0.163    \\ 
\textit{\textbf{\begin{tabular}[c]{@{}c@{}}CWT-C-VAWGAN\end{tabular}}} & 2.808 $\pm$ 0.137    \\ \hline
\end{tabular}
\vspace{-3mm}
\label{tab:mos}
\end{table}

\begin{figure}[ht]
    \vspace{-3mm}
    \centering
    \includegraphics[width=6cm]{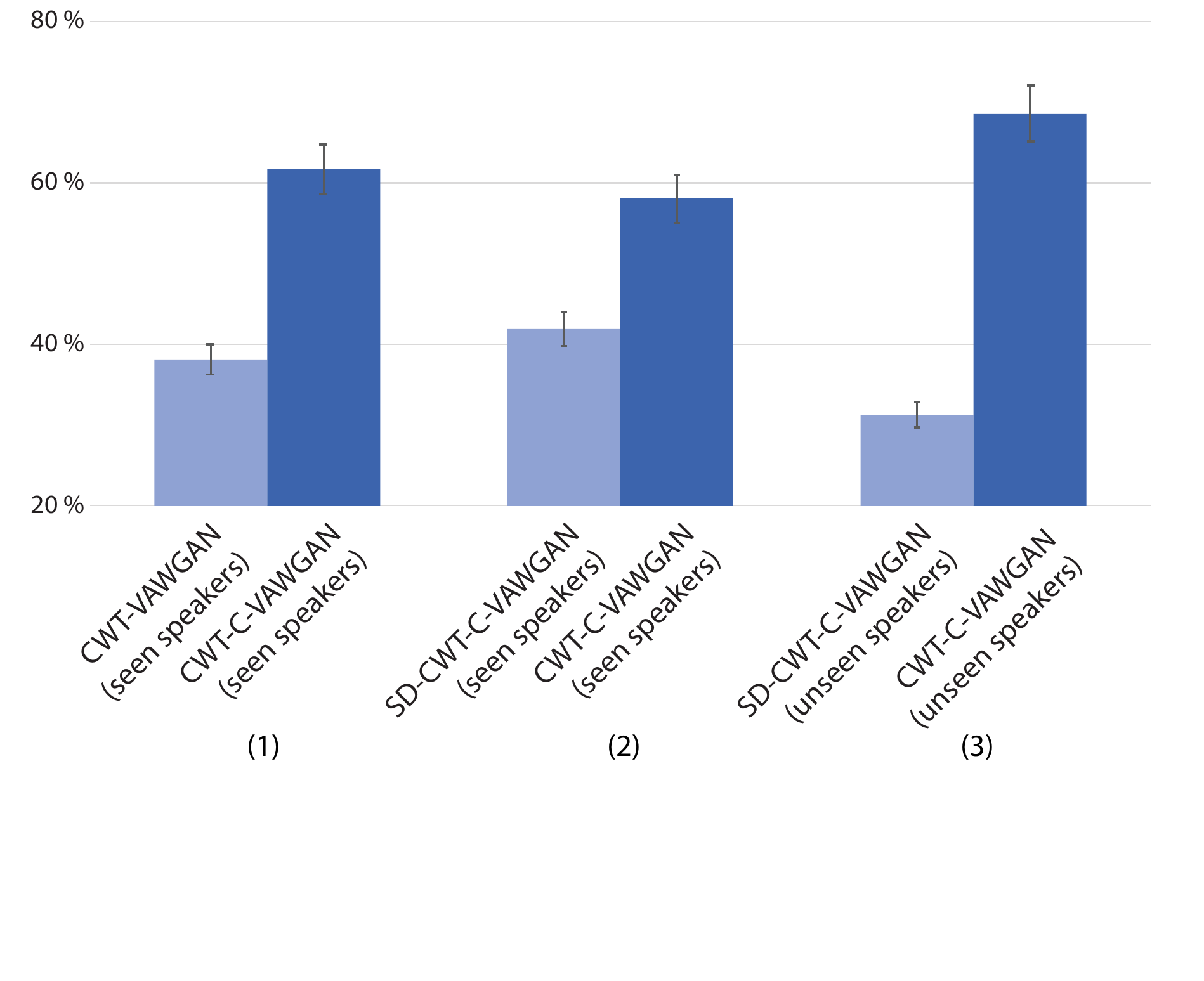}
    \vspace{-14mm}
    \caption{XAB emotion similarity preference results with 95\% confidence interval to assess: (1) the effect of F0 conditioning for seen speakers, (2) speaker-dependent vs speaker-independent training for seen speakers, and (3) speaker-dependent vs speaker-independent training for unseen speakers. }
    \label{fig:ab1}
    \vspace{-2mm}
\end{figure}

We further conduct XAB emotion similarity test to assess the emotion conversion performance, where the subjects are asked to choose the speech sample which sounds closer to the reference in terms of emotional expression. As shown in Fig. \ref{fig:ab1}(1), we observe that \textit{CWT-C-VAWGAN} clearly outperforms \textit{CWT-VAWGAN} in terms of emotion similarity for seen speakers. It shows that conditioning on the converted CWT-F0 features further improves the emotional expression. 

We also conduct XAB emotion similarity test to assess the performance of proposed framework between speaker-independent \textit{CWT-C-VAWGAN} and speaker-dependent training \textit{SD-CWT-C-VAWGAN}, as reported in Fig.\ref{fig:ab1}(2) for seen speakers, and in Fig.\ref{fig:ab1}(3) for unseen speakers.  \textit{SD-CWT-C-VAWGAN} is trained with data only from one specific speaker. We train the baseline separately for each of the three specific speakers and perform speaker-dependent tests. We observe that speaker-independent training outperforms speaker-dependent training for both seen and unseen speakers, that is very encouraging. We think that speaker-independent training benefits from a multi-speaker database and learns the speaker-independent emotion mapping effectively. We also observe that the listeners strongly favor speaker-independent training over speaker-dependent training for unseen speakers.



 

\begin{figure}[ht]
    \vspace{-4mm}
    \centering
    \includegraphics[width=5cm]{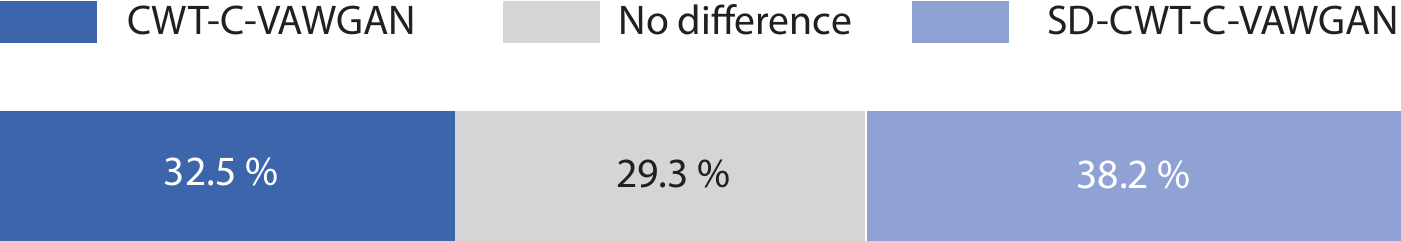}
    \vspace{-2mm}
    \caption{XAB speaker similarity preference results of speaker-dependent and speaker-independent training for seen speakers. }
    \label{fig:ab2}
\end{figure}

Lastly, we conduct XAB speaker similarity test to compare the  performance of the proposed speaker-independent training \textit{CWT-C-VAWGAN} and speaker-dependent training \textit{SD-CWT-C-VAWGAN}. We note that \textit{SD-CWT-C-VAWGAN} is trained only with the specific speaker, hence expected to have a better performance in speaker similarity. As reported in Fig.\ref{fig:ab2}, we observe that \textit{CWT-C-VAWGAN} achieves comparable results with \textit{SD-CWT-C-VAWGAN} that we believe is an encouraging outcome. The results indicate that \textit{CWT-C-VAWGAN} framework does not convert the emotion at the expense of speaker similarity and shows remarkable performance of preserving the speaker identity while performing speaker-independent emotion conversion.
\vspace{-8mm}
\subsection{Discussion}
\vspace{-1mm}
The experiments suggest that: (1) The proposed framework learns the speaker-independent emotional expression pattern for both spectrum and prosody across speakers; (2) The speaker-independent training outperforms speaker-dependent training, while successfully preserving speaker identity of the source speaker; (3) The proposed CWT-based F0 conditioning improves spectrum conversion;  and (4) The proposed framework is capable of converting anyone's emotion. To our best knowledge, this paper is the first to provide a speaker-independent perspective to emotion conversion.  



\vspace{-5mm}
\section{Conclusions}
We propose a speaker-independent EVC framework that converts anyone's emotion without the need for parallel data. We perform both spectrum and prosody conversion based on VAW-GAN. We provide CWT modelling of F0 to describe the prosody in different time resolutions. Moreover, we study the use of CWT-based F0 as an additional input to the decoder to improve the spectrum conversion performance. Experimental results validate the idea of speaker-independent EVC by showing remarkable performance for both seen and unseen speakers. 
\vspace{-6mm}
\section{Acknowledgement}
This work is supported by National Research Foundation Singapore under the AI Singapore Programme (Award Number: AISG-100E-2018-006, AISG-GC-2019-002), under the National Robotics Programme (Grant Number: 192 2500054), and Programmatic Grant No. A18A2b0046 (Human Robot Collaborative AI for AME) and A1687b0033 (Neuromorphic Computing) from the Singapore Government’s Research, Innovation and Enterprise 2020 plan in the Advanced Manufacturing and Engineering domain.
This work is also supported by SUTD Start-up Grant Artificial Intelligence for Human Voice Conversion (SRG ISTD2020 158) and SUTD AI Grant titled ’The Understanding and Synthesis of Expressive Speech by AI’ (PIE-SGP-AI-2020-02).

\bibliographystyle{IEEEtran}
{\footnotesize
\bibliography{mybib}
}
\end{document}